\documentclass[prd,twocolumn,a4paper,superscriptaddress,nofootinbib]{revtex4}
\usepackage{amsmath,amssymb,epsfig,bm}

\begin{document}

%%%%%%%%%%%%%%%%%%%%%%%%%%%%%%%%%%%%%%%%%%%%%%%%%%%%%%%%%%%%%%%%%%
%                       Own commands                             %
%%%%%%%%%%%%%%%%%%%%%%%%%%%%%%%%%%%%%%%%%%%%%%%%%%%%%%%%%%%%%%%%%%

\newcommand{\vnhat}{\hat{\mathbf{n}}}
\newcommand{\valpha}{\bm{\alpha}}
\newcommand{\edth}{\,\eth\,}
\newcommand{\bedth}{\,\overline{\eth}\,}

%%%%%%%%%%%%%%%%%%%%%%%%%%%%%%%%%%%%%%%%%%%%%%%%%%%%%%%%%%%%%%%%%%
%                       Title etc.                               %
%%%%%%%%%%%%%%%%%%%%%%%%%%%%%%%%%%%%%%%%%%%%%%%%%%%%%%%%%%%%%%%%%%

\title{Geometry of weak lensing of CMB polarization}

\author{Anthony Challinor}
\email{A.D.Challinor@mrao.cam.ac.uk}
\author{Gayoung Chon}
\email{gchon@mrao.cam.ac.uk}
 \affiliation{Astrophysics Group, Cavendish Laboratory, Madingley Road,
Cambridge CB3 OHE, United Kingdom}

\begin{abstract}
Hu has presented a harmonic-space method for calculating the effects
of weak gravitational lensing on the cosmic microwave background (CMB) over
the full sky. Computing the lensed power spectra to first order in the
deflection power requires one to formulate the lensing displacement
beyond the tangent-space approximation. We point out that for CMB polarization
this displacement must undergo geometric corrections on the spherical sky to
maintain statistical isotropy of the lensed fields. Although not discussed
by Hu, these geometric effects are implicit in his analysis. However, there
they are hidden by an overly-compact notation that is both unconventional and
rather confusing. Here we aim to ameliorate this deficiency by providing
a rigorous derivation of the lensed spherical power spectra.
\end{abstract}


\maketitle

%%%%%%%%%%%%%%%%%%%%%%%%%%%%%%%%%%%%%%%%%%%%%%%%%%%%%%%%%%%%%%%%%%
%                         Main body                              %
%%%%%%%%%%%%%%%%%%%%%%%%%%%%%%%%%%%%%%%%%%%%%%%%%%%%%%%%%%%%%%%%%%

The re-mapping of temperature and polarization anisotropies by weak
gravitational lensing can lead to observable distortions in the
cosmic microwave background (CMB). In Ref.~\cite{hu2000}, Hu presented
an elegant harmonic-space method for calculating second and higher-order
statistics of the lensed CMB temperature and polarization fields beyond the
flat-sky approximation. Hu showed that to compute the lensing effects on the
CMB accurately it is necessary to use the full-sky power spectrum of the
lensing deflection potential. Simple expressions for the lensed power spectra
in an all-sky formalism were derived, and their use was advocated for analysis
work.

Although we
agree with the final results presented in Ref.~\cite{hu2000} for the lensed
power spectra, we regard several steps in the published derivation as
rather confusing. In particular, the re-mapping of the primary polarization
on the sphere caused by gravitational lensing is only implicit in Hu's
analysis [although it is given explicitly for the flat-sky case
in his Eq.~(43)]. The point we wish to emphasise here is that on the sphere
it is no longer appropriate simply to displace the Stokes parameters (defined
on the polar coordinate basis) by the deflection vector; this
operation is not covariant and would not maintain statistical isotropy of the
lensed fields. This potential confusion is not helped
by the overly-compact, and unconventional, derivative notation that Hu has
introduced in Ref.~\cite{hu2000} without comment. For example, in his Eq.~(65)
if the derivatives acting on spin-weight 2 objects are interpreted as usual
covariant derivatives on the sphere (as suggested by the notation $\nabla_i$), then the lens re-mapping one infers from his Eq.~(64) is a simple
displacement of Stokes parameters in the polar basis. Here we develop the
appropriate re-mapping to describe lensing of the polarization field on the
spherical sky in both the tensor and spin-weight formalisms.
We then proceed to give a rigorous derivation of the all-sky lensed
polarization power spectra that we hope will eliminate any further confusion
that may arise from the presentation in Ref.~\cite{hu2000}.

In the Born approximation, the lensing deflection is calculated on the
unlensed line of sight so the lensing map is a local function of the
deflection vector $\alpha^a = \nabla^a \psi$, where $\psi(\vnhat)$ is the
lensing potential (e.g.\ Ref.~\cite{kaiser98}).
Even for a scalar field, such as
the CMB temperature anisotropy $\Theta(\vnhat)$, there is some ambiguity
in how to interpret the deflection vector on the spherical sky. Expressions
like $\tilde{\Theta}(\vnhat) = \Theta(\vnhat + \valpha)$, relating the unlensed
temperature $\Theta(\vnhat)$ to the lensed $\tilde{\Theta}(\vnhat)$, are
only unambiguous in the tangent-space approximation: $\tilde{\Theta}(\vnhat)
= \Theta(\vnhat) + \alpha^a \nabla_a \Theta(\vnhat)$. However, to calculate the
lensed power spectra to first order in the deflection power we need
the mapping to second order in $\alpha^a$. The choice made in
Ref.~\cite{hu2000}, albeit without comment, is to take
\begin{equation}
\tilde{\Theta}(\vnhat)
= \Theta(\vnhat) + \alpha^a \nabla_a \Theta(\vnhat) + \tfrac{1}{2}
\ \alpha^a \alpha^b \nabla_a \nabla_b \Theta(\vnhat) + \cdots .
\label{eq:1}
\end{equation}
The geometric interpretation of this mapping is that one should displace from
$\vnhat$ for geodesic distance $|\alpha^a(\vnhat)|$ along the geodesic through
$\vnhat$ that has initial unit tangent vector $\hat{\alpha}^a(\vnhat)$.
The mapping
in Eq.~(\ref{eq:1}) has the desirable property that it is local in $\alpha^a$
(i.e.\ the lensed field at $\vnhat$ only depends on $\alpha^a$ at that point),
consistent with the Born approximation.

For polarization, simply displacing the Stokes parameters according to
Eq.~(\ref{eq:1}) is a coordinate-dependent
operation and hence unphysical. This would show up in spurious correlations
between the multipoles of the lensed polarization field that should vanish
for a statistically-isotropic field. The appropriate mapping for CMB
polarization involves parallel transporting the linear polarization tensor
$\mathcal{P}_{ab}$ (whose components in a
specific orthonormal basis are the Stokes parameters) back along the geodesic
generated at $\vnhat$ by the unit tangent $\hat{\alpha}^a(\vnhat)$. We can only
apply Eq.~(\ref{eq:1}) directly to the Stokes parameters if they are defined
on basis vectors parallel and perpendicular to the geodesic at
$\vnhat$ and its image point.

To establish the appropriate analogue of Eq.~(\ref{eq:1}) for polarization,
we define a tensor $\Delta_{ab}(s) \equiv \mathcal{P}_{ab}[x(s)] -
\mathcal{P}_{\| ab}(s)$ along the geodesic $x^a(s)$ connecting
$\vnhat$ and its image. Here, $s$ is distance along the geodesic
measured from $\vnhat$, so $s=s_* \equiv |\alpha^a(\vnhat)|$ at the end point,
and $\mathcal{P}_{\| ab}(s)$ is the tensor obtained by parallel transporting
$\mathcal{P}_{ab}[x(s_*)]$ back along the geodesic. With these definitions,
$\Delta_{ab}(s_*) = 0$ and the lensed polarization tensor at $\vnhat$ is
$\tilde{\mathcal{P}}_{ab}(\vnhat) = \mathcal{P}_{ab}(\vnhat) - \Delta_{ab}(0)$.
We now compute $\Delta_{ab}(s_*)$ by Taylor expanding its components
$\Delta_{ij}$ in an arbitrary basis about $s=0$. Setting
the expansion to zero and using the geodesic character of $x^a(s)$,
to second order we find
\begin{equation}
\Delta_{ij}(0) = - \alpha^k \partial_k \Delta_{ij} + \tfrac{1}{2}
\alpha^l \alpha^m \left(\Gamma^k_{lm} \partial_k \Delta_{ij} -
\partial_l \partial_m \Delta_{ij}\right), 
\label{eq:2}
\end{equation}
where all terms on the right are evaluated at $s=0$. Replacing the
partial derivatives with covariant derivatives and terms involving the
connection $\Gamma^k_{lm}$, and solving iteratively for $\Delta_{ab}(0)$,
we find
\begin{equation}
\Delta_{ab}(0) = - \alpha^c \nabla_c \Delta_{ab} - \tfrac{1}{2}
\alpha^c \alpha^d \nabla_c \nabla_d \Delta_{ab} + \cdots.
\label{eq:3}
\end{equation}
Finally, we use the fact that $\mathcal{P}_{\| ab}(s)$ is parallel
transported along $x^a(s)$ to obtain our required result for the
lensed polarization at $\vnhat$:
\begin{equation}
\tilde{\mathcal{P}}_{ab}(\vnhat) = \mathcal{P}_{ab}(\vnhat)
+ \alpha^c \nabla_c \mathcal{P}_{ab}(\vnhat) + \tfrac{1}{2}
\alpha^c \alpha^d \nabla_c \nabla_d \mathcal{P}_{ab}(\vnhat) + \cdots. 
\label{eq:4}
\end{equation}
This expansion is the obvious generalisation of Eq.~(\ref{eq:1}) to the
covariant transport of polarization.

The expansion of the symmetric, trace-free polarization tensor in the
spin-weight formalism~\cite{goldberg67}
(our conventions follow Ref.~\cite{lewis02}) is
\begin{equation}
\mathcal{P}^{ab} = 2^{-2} {}_{-2}P e_+^a e_+^b + 2^{-2} {}_{2} P
e_-^a e_-^b,
\label{eq:5}
\end{equation}
where it is convenient to choose the complex null vectors
$e_{\pm}^a = (\partial_\theta)^a \pm i \csc \theta (\partial_\phi)^a$.
With this choice, the spin $\pm 2$ polarization ${}_{\pm 2} P = Q \mp i U$,
where the Stokes
parameters $Q$ and $U$ are measured on the $\{(\partial_\theta)^a,
- \csc\theta (\partial_\phi)^a\}$ basis. With a little effort,
we can establish the spin-weight version of Eq.~(\ref{eq:4}):
\begin{eqnarray}
{}_{-2} \tilde{P} &=& {}_{-2}P - \tfrac{1}{2} ({}_1 \alpha \bedth +
{}_{-1}\alpha \edth) {}_{-2}P \nonumber \\
&&\mbox{} + \tfrac{1}{8} ({}_1 \alpha {}_1 \alpha
\bedth \bedth + {}_1 \alpha {}_{-1}\alpha \edth\bedth \nonumber \\
&&\mbox{}\phantom{+ \tfrac{1}{8} (}+ {}_{-1}\alpha
{}_1 \alpha \bedth \edth + {}_{-1} \alpha {}_{-1} \alpha \edth\edth)
{}_{-2} P + \cdots , 
\label{eq:6}
\end{eqnarray}
where ${}_{\pm 1}\alpha$ are the spin-weight components of $\alpha^a$:
$\alpha^a = 2^{-1} {}_1 \alpha e_-^a + 2^{-1} {}_{-1}\alpha e_+^a$.
In terms of the
raising and lowering operators~\cite{goldberg67,lewis02}, $\edth$ and $\bedth$,
and the lensing potential $\psi(\vnhat)$, 
we have ${}_1 \alpha = - \edth \psi$ and ${}_{-1} \alpha = {}_1 \alpha^\ast
= - \bedth \psi$. Equation~(\ref{eq:6}), which holds quite generally for the
spin-weight components ${}_{\pm s}\eta$ of any symmetric, trace-free
tensor $\eta_{a_1 \dots a_s}$, is the correct covariant generalisation of the
flat-sky result given as Eq.~(43) in Ref.~\cite{hu2000}. The
difference between Eq.~(\ref{eq:6}) and a simple Taylor expansion of
${}_{-2} P$, i.e.\ ${}_{-2} \tilde{P} = {}_{-2} P + \alpha^a \nabla_a
{}_{-2} P + \cdots$, amounts to a factor $e^{-2i\chi}$. This factor describes
the combined effect of the rotations at $\vnhat$ and its image that are
required to align one of the polarization basis vectors with the local tangent
to the geodesic. The angle $\chi$ has the expansion
\begin{eqnarray}
\chi &=& \tfrac{1}{2} i e_-^a \alpha^b \nabla_b e_{+a}
- \tfrac{1}{8} i (e_-^a \alpha^b \nabla_b e_{+a})^2 \nonumber \\
&&\mbox{} + \tfrac{1}{4}i
e_-^a \alpha^b \alpha^c \nabla_b \nabla_c e_{+a} + \cdots.
\label{eq:7}
\end{eqnarray}
Evaluating the covariant derivatives gives $\chi$ in terms of the (polar)
coordinate components of $\alpha^a$:
\begin{equation}
\chi = \alpha^\phi \left[ \cos\theta - \tfrac{1}{2} \csc\theta(1+\cos^2\theta)
\alpha^\theta + \cdots\right]. 
\label{eq:8}
\end{equation}
When displacing along lines of constant $\phi$ the rotation angle vanishes,
as expected.

Having established the appropriate form of the polarization mapping describing
weak lensing on the sphere, it is straightforward to compute the power
spectra of the lensed fields. One can either work directly with the
polarization tensor, using Eq.~(\ref{eq:5}) and an expansion in the tensor
harmonics~\cite{kamionkowski97}, or proceed in the spin-weight formalism
using Eq.~(\ref{eq:6}) and an expansion in spin-weight
harmonics~\cite{seljak97}. Here we shall follow the latter route and sketch
a rigorous derivation of the polarization power spectra. Extracting the
spin-$\pm 2$ components, ${}_{\pm 2}\tilde{P}_{lm}$, of the lensed polarization
using Eq.~(\ref{eq:6}) we find
\begin{eqnarray}
{}_{\pm 2}\tilde{P}_{lm} &=& {}_{\pm 2}P_{lm} + \psi_{(lm)_1}
{}_{\pm 2}P_{(lm)_2} {}_{\pm 2}I_{ll_1 l_2}^{m m_1 m_2} \nonumber \\
&&\mbox{}
+ \tfrac{1}{2} \psi_{(lm)_1} \psi^\ast_{(lm)_2} {}_{\pm 2}P_{(lm)_3}
{}_{\pm 2}J_{ll_1 l_2 l_3}^{m m_1 m_2 m_3},
\label{eq:9}
\end{eqnarray}
where summation should be understood over repeated multipole indices ($l$
and $m$). 
Here ${}_{\pm 2}P_{lm}$ are the spin-weight multipoles of the unlensed
polarization and $\psi_{lm}$ are the usual spherical multipoles of the scalar
lensing potential. The overlap integrals are
\begin{eqnarray}
{}_{\pm 2} I_{ll_1 l_2}^{mm_1 m_2} &\equiv& \int d\vnhat \,
{}_{\pm 2}Y_{lm}^\ast \tfrac{1}{2}
(\edth Y_{(lm)_1} \bedth + \bedth Y_{(lm)_1} \edth) \nonumber \\
&&\mbox{} \phantom{\frac{1}{2} \int d\vnhat \,
{}_{\pm 2}Y_{lm}^\ast (}
\times {}_{\pm 2} Y_{(lm)_2}, \label{eq:10} \\
{}_{\pm 2} J_{ll_1 l_2 l_3}^{mm_1 m_2 m_3} &\equiv& \int d\vnhat \,
{}_{\pm 2}Y_{lm}^\ast \tfrac{1}{4}[\edth Y_{(lm)_1} \edth Y_{(lm)_2}^\ast
\bedth\bedth \nonumber \\
&&\mbox{} + \bedth Y_{(lm)_1} \bedth Y_{(lm)_2}^\ast \edth\edth \nonumber \\
&&\mbox{} + \edth Y_{(lm)_1} \bedth Y_{(lm)_2}^\ast
(\bedth \edth + \edth \bedth)] \nonumber \\
&&\mbox{} \times {}_{\pm 2} Y_{(lm)_3},
\label{eq:11}
\end{eqnarray}
which should be compared with Eq.~(65) in Ref.~\cite{hu2000}\footnote{%
There is a typo in the second line in Eq.~(65) of Ref.~\cite{hu2000}: one
of the derivatives acting on the spin-0 harmonics should be $\nabla_j$
rather than $\nabla_i$.}.
The integral defining ${}_{\pm 2} I_{ll_1 l_2}^{mm_1 m_2}$ can be evaluated by
integrating by parts using the spin-weight integral theorems in the appendices
of Ref.~\cite{lewis02}, and noting that the spin-weight harmonics ${}_s Y_{lm}$
are eigenfunctions of $\edth \bedth$ with eigenvalue $s(s-1)-l(l-1)$. [The
identity $(\bedth\edth - \edth\bedth){}_s \eta = 2 s {}_s \eta$ for spin-$s$
${}_s \eta$ is also useful.] The integral for
${}_{\pm 2} I_{ll_1 l_2}^{mm_1 m_2}$ then reduces to the right-hand side
of Eq.~(72) in Ref.~\cite{hu2000}, which can be evaluated in terms of
3-$j$ symbols. 

Breaking the spin-weight multipoles into parity eigenstates,
${}_{\pm 2}P_{lm} = E_{lm} \pm i B_{lm}$, where $E_{lm}$ are the multipoles
of electric polarization and $B_{lm}$ of magnetic, and assuming no correlations
between polarization and the lensing potential, the overlap integrals
${}_{\pm 2} I$ enter the correlators such as
$\langle \tilde{E}_{lm} \tilde{E}_{(lm)'}^\ast \rangle$ in the form
$\sum_{m_1 m_2} {}_2 I_{ll_1 l_2}^{m m_1 m_2} {}_2 I_{l' l_1 l_2}^{m' m_1 m_2}
{}^\ast$. The orthogonality relations for the 3-$j$ symbols ensure that the
summation vanishes unless $l=l'$ and $m=m'$ as required to maintain statistical
isotropy of the lensed fields. The overlap integrals
${}_{\pm 2} J$ enter the correlators in the form
$\sum_{m_1} {}_{\pm 2} J_{ll_1 l_1 l'}^{m m_1 m_1 m'}$ (and its complex
conjugate). Setting $m_1 = m_2$ in Eq.~(\ref{eq:11}), and summing, isolates
terms
\begin{equation}
\sum_{m_1} \edth Y_{(lm)_1}\bedth Y_{(lm)_1}^\ast = l_1 (l_1+1)
\frac{2l_1+1}{4\pi},
\label{eq:12}
\end{equation}
and $\sum_{m_1} \edth Y_{(lm)_1}\edth Y_{(lm)_1}^\ast = 0$, where we have
used the addition theorem for spin-weight harmonics. It follows that only
the term involving the operator $\bedth \edth + \edth \bedth$ in
${}_{\pm 2} J$ survives the summation. Finally, making use of
\begin{equation}
\tfrac{1}{2} (\bedth \edth + \edth \bedth) {}_s Y_{lm} = [s^2-l(l+1)]
{}_s Y_{lm},
\label{eq:13}
\end{equation}
we find
\begin{equation}
\sum_{m_1} {}_{\pm 2} J_{ll_1 l_1 l'}^{m m_1 m_1 m'} = - \frac{1}{2}
[l(l+1)-4]l_1 (l_1+1)\frac{2l_1 + 1}{4\pi} \delta_{ll'} \delta_{mm'},
\label{eq:14}
\end{equation}
as required to maintain statistical isotropy. Putting these results together
gives the final expressions for the lensed power spectra which agree with
those given by Hu as Eq.~(76) in Ref.~\cite{hu2000}. Our result for the
cross power spectrum between $E$ and $\Theta$ also agrees with
Ref.~\cite{hu2000}.

We close by discussing the relation of our derivation to that in
Ref.~\cite{hu2000}. The critical point to note is that if the
covariant derivatives there, acting on a spin-weight $s$ ($\geq 0$) quantity
${}_s \eta$, are interpreted as encoding the operation $2^{-s} e_+^{A_s}
\nabla_b ({}_s \eta e_{-A_s})$, where $e_\pm^{A_s}$ denotes the irreducible
tensor product $e_\pm^{a_1} \dots e_\pm^{a_s}$, then the lens mapping
implied by Eq.~(64) of Ref.~\cite{hu2000} reduces to that given here.
(The result that $e_+^{A_s} e_{- B_s}$ is covariantly
constant proves useful.)
Furthermore, it can be shown that the integration by parts
used to simplify the overlap integrals ${}_{\pm 2}I$ remains consistent under
this interpretation of the covariant derivative acting on spin-weight
quantities. Unfortunately, this compressed notation appears to have been
adopted in Ref.~\cite{hu2000} without comment. [It is hinted at in
Eq.~(55) there, and in the comment above Eq.~(71).] Since the 
raising and lowering operators of the spin-weight formalism naturally
encode the required derivative operations, there is no need to
overload the standard notation of differential geometry. Furthermore,
the overly-compact notation of Ref.~\cite{hu2000} hides the covariant nature
of the lens re-mapping of polarization on the sphere.

%%%%%%%%%%%%%%%%%%%%%%%%%%%%%%%%%%%%%%%%%%%%%%%%%%%%%%%%%%%%%%%%%%
%                       Acknowledgments                          %
%%%%%%%%%%%%%%%%%%%%%%%%%%%%%%%%%%%%%%%%%%%%%%%%%%%%%%%%%%%%%%%%%%

\begin{acknowledgments}
AC acknowledges a PPARC Postdoctoral Fellowship. We thank Wayne Hu for
useful comments.
\end{acknowledgments}

%%%%%%%%%%%%%%%%%%%%%%%%%%%%%%%%%%%%%%%%%%%%%%%%%%%%%%%%%%%%%%%%%%
%                       References                               %
%%%%%%%%%%%%%%%%%%%%%%%%%%%%%%%%%%%%%%%%%%%%%%%%%%%%%%%%%%%%%%%%%%

%\bibliography{CC02}

\end{document}